\documentclass[manuscript]{acmart}
\AtBeginDocument{%
  }

\setcopyright{acmlicensed}
\copyrightyear{2025}
\acmYear{2025}
\acmDOI{10.1145/3706599.3719897}
\acmConference[CHI '25]{In
Proceedings of the CHI Conference on Human Factors in Computing Systems}{Apr 26--May 01, 2025}{Yokohama, Japan}
\acmISBN{978-1-4503-XXXX-X/2018/06}

\begin{document}

\title{StreetScape: Gamified Tactile Interactions for Collaborative Learning and Play}

 \author{Areen Khalaila}
\email{areenkh@brandeis.edu}
\affiliation{%
  \institution{Brandeis University}
  \city{Waltham}
  \state{Massachusetts}
  \country{USA}
}

\author{Gianna Everette}
\affiliation{%
  \institution{Brandeis University}
  \city{Waltham}
    \state{Massachusetts}
  \country{USA}}
\email{gleverette@brandeis.edu}

\author{Suho Kim}
\email{kim804@brandeis.edu}
\affiliation{%
  \institution{Brandeis University}
  \city{Waltham}
    \state{Massachusetts}
  \country{USA}
}

\author{Ian Roy}
\email{ianroy@brandeis.edu}
\affiliation{%
  \institution{Brandeis University}
  \city{Waltham}
    \state{Massachusetts}
  \country{USA}
}

\email{anon@example.com}



\begin{abstract}
Spatial reasoning and collaboration are essential for childhood development, yet blind and visually impaired (BVI) children often lack access to tools that foster these skills. Tactile maps and assistive technologies primarily focus on individual navigation, overlooking the need for playful, inclusive, and collaborative interactions. We address this with StreetScape, a tactile street puzzle that enhances spatial skills and interdependence between BVI and sighted children. Featuring modular 3D-printed tiles, tactile roadways, and customizable decorative elements, StreetScape allows users to construct and explore cityscapes through gamified tactile interaction.  Developed through an iterative design process, it integrates dynamic assembly and tactile markers for intuitive navigation, promoting spatial learning and fostering meaningful social connections. This work advances accessible design by demonstrating how tactile tools can effectively bridge educational and social gaps through collaborative play, redefining assistive technologies for children as a scalable platform that merges learning, creativity, and inclusivity.
\end{abstract}





\maketitle

\section{Introduction}
Blind and visually impaired (BVI) children face unique challenges in developing spatial awareness and navigating physical environments, which are critical for independence and collaboration \cite{Wei2022, Brule2016}. 
Assistive technologies and tactile maps have been extensively developed to support navigation and spatial skills for BVI adults
(e.g., \cite{Hofmann2022, Crawford2024, Manuel2023, Kuribayashi2021, Kupferstein2020, Kuribayashi2024, Guerreiro2019, Kayukawa2023}). However, interventions tailored specifically for BVI children are sparse, despite their global population reaching 19 million \cite{WHO2015} and the critical importance of early specialized intervention in their developmental trajectories \cite{Fielder1993}. Additionally, existing resources often fail to integrate inclusivity and interdependence—key principles that empower people with disabilities to engage meaningfully with their  their non-disabled peers \cite{Bennett2018}. Furthermore, many of tools prioritize academic or navigational outcomes, overlooking the potential for gamified, collaborative designs that promote both play and learning. Tactile maps have been widely studied for their role in supporting wayfinding and spatial orientation for BVI individuals \cite{Jacobson1992, cogprints1509}, with research highlighting both the advantages and limitations of 3D-printed tactile maps \cite{Hofmann2022, Jido2019, Taylor2016}. However, their potential as gamified, collaborative tools designed specifically to support spatial learning and social interaction among children remains largely unexplored. Gamification refers to the integration of game-like elements—such as rewards, challenges, and interactive feedback—into non-game contexts to enhance motivation and engagement \cite{gamification}.
 In educational settings, gamification has been shown to enhance engagement and learning outcomes \cite{hussein2023gamification, Kiryakova}, particularly for children with disabilities, by making complex tasks more accessible and motivating \cite{Cahyono2023, mahmoudi2024gamification}. Integrating gamification into tactile maps could foster interactive learning experiences, encourage peer collaboration, and improve spatial reasoning skills in young BVI learners.

To address this gap, we developed \textit{StreetScape}, a tactile street map puzzle designed to foster spatial learning and collaborative play among BVI and sighted children. Drawing inspiration from prior work advocating for interdependence, which emphasizes the value of mutual learning experiences for disabled and non-disabled individuals \cite{Bennett2018}, StreetScape introduces an interactive platform where children collaboratively construct a tactile representation of a cityscape. The puzzle consists of modular 3D-printed components, including street tiles, decorative elements such as streetlamps and stoplights, and a gridded baseboard with indented slots for precise tile placement. The map is designed with tactile accessibility as a priority, featuring raised textures and varying shapes to ensure each component is easily distinguishable by touch. The street tiles represent different configurations, such as intersections, curves, and straight roads, while the decorative elements are designed to be placed at corresponding slots to enhance spatial understanding. The iterative development process of StreetScape focused on refining its tactile clarity, durability, and ease of assembly. Materials such as polylactic acid (PLA), polyethylene terephthalate glycol (PETG), and carbon fiber-based filament were tested to ensure a balance between robustness and usability. The modular design of the puzzle allows for flexibility in creating different layouts, making it a customizable and scalable tool. These design considerations result in a cohesive tactile experience that aligns with the needs of BVI children, ensuring precise and intuitive interaction with the map.

\section{Related Work}

\subsection{Assistive Technologies for BVI children}
Assistive technologies have long aimed to foster independence for children with disabilities \cite{Gadiraju2024, Ruth1959, swallow1987thrive}, providing tools that enable them to navigate their environments \cite{Brule2016, Jaime2010, Freeman2017}, learn \cite{Mikulowski2019, AudioMath, StepSure}, and develop essential skills \cite{Manuel2023, Rocha2021}. 
Tools such as LucentMaps, a 3D-printed audiovisual tactile map system designed for BVI users, integrate capacitive touch-sensitive elements with mobile devices to provide interactive auditory feedback, enhancing spatial learning and navigation\cite{LucentMaps}. Similarly, Jido combine interactive tactile maps with auditory feedback to improve spatial orientation and learning experiences \cite{jido}. These systems leverage real-time audio guidance to compensate for the lack of visual information, allowing BVI users to construct mental maps of their surroundings.
While these tools successfully enhance accessibility, they primarily focus on individual learning and navigation, rather than collaborative or gamified interactions. StreetScape differentiates itself by incorporating interdependence and social engagement into the learning process, allowing BVI and sighted children to build and explore spatial layouts together. Unlike audio-tactile maps that provide structured, pre-programmed feedback, StreetScape promotes exploratory learning and open-ended play, fostering deeper spatial reasoning skills and collaborative problem-solving. By positioning StreetScape within this landscape, we highlight the need for assistive technologies that extend beyond navigation aids and engage users in interactive, cooperative spatial learning.

Recent research in assistive technology design has begun shifting from independence as the sole goal to embracing interdependence—a frame that emphasizes the collaborative creation of access and the shared contributions of all users \cite{Bennett2018}. This approach is particularly important for fostering interactions between BVI and sighted children, enabling them to co-create inclusive experiences. For example, tools like Incloodle have demonstrated how assistive technologies can facilitate equal participation among neurodiverse and neurotypical children in shared activities \cite{Sobel2016}.

While prior studies have highlighted the value of collaboration between BVI children and their sighted parents in skill development \cite{Gadiraju2024}, the importance of peer interactions remains underexplored. Peer interactions can provide unique opportunities for BVI children to engage in age-appropriate social exchanges, develop a sense of belonging, and learn from diverse perspectives—a crucial component of childhood development \cite{holbrook2000foundations}. Tools that promote collaborative play and learning between BVI and sighted children have the potential of bridging this gap, fostering interdependence and mutual understanding while preparing children for inclusive environments.

\subsection{Tactile Maps for BVI users}
Tactile maps have been extensively studied as tools for supporting navigation and spatial orientation among BVI adults. These maps offer an accessible medium for conveying spatial relationships and geographic information, enabling users to plan routes and familiarize themselves with unfamiliar environments \cite{cogprints1509, Jacobson1992, Crawford2024}. Prior work has explored the integration of tactile features with additional modalities, such as audio feedback, to enhance usability and reduce reliance on Braille, which many BVI individuals do not read \cite{Coughlan2022, CamIO2013}. These studies highlight the importance of tailoring tactile maps to the diverse needs and preferences of users as well as adapting maps for different contexts, including indoor and outdoor environments \cite{Williams2013, Gupta2020}.

Despite significant advancements, much of the existing research focuses on adults and their specific navigational needs, with limited attention given to tactile map designs tailored for BVI children. Unlike adults, children require tools that not only convey spatial information but also integrate playfulness and interactivity to foster engagement and learning \cite{pyle2017play, yogman2018power}. Furthermore, while tactile maps for adults often prioritize efficiency and practicality, children’s tools must consider developmental goals, such as building spatial reasoning and collaborative skills. The lack of research addressing these nuanced requirements underscores the need for tactile tools designed specifically for BVI children, which incorporate elements like gamification and encourage interdependent interactions with sighted peers.

\section{StreetScape}

\subsection{Core Design Concepts}
StreetScape is conceptualized to support BVI children between the ages of six and ten. This target age range was chosen based on developmental milestones related to spatial reasoning, motor skills, and social interaction.  This age group is in the concrete operational stage of cognitive development, during which they begin to think logically and understand spatial relationships more effectively. \footnote{\url{https://www.cincinnatichildrens.org/health/c/cognitive-development}} Additionally, this age group has the fine motor skills needed to manipulate tactile puzzle pieces and assemble the map with precision, a skill that typically begins to stabilize around age six. \footnote{\url{https://www.columbiadoctors.org/health-library/special/growth-development-ages-6-10-years/}}  By focusing on this age group, StreetScape seeks to balance cognitive, social, and motor skill development with the playfulness and accessibility required to engage young learners. Through our collaboration with three teachers from The Carroll Center for the Blind, all of whom identify as having low vision, we found two critical gaps. First, many BVI children tend to be more cautious and hesitant when engaging with others or exploring their surroundings due to limited spatial awareness. Without a clear mental representation of the spaces they inhabit, children often experience the world through fragmented tactile encounters, which can restrict their understanding \cite{landau1984spatial}. For example, while a sighted child may recognize a car as a cohesive structure with four wheels, doors, and windows, a BVI child may only associate it with isolated tactile features, such as a seat or a door handle. 

Second, there is a significant need for inclusively designed educational tools and games. Our collaborators at The Carroll Center, specialists in blindness education and assistive technology, provided critical insights into the needs of BVI children during spatial learning activities. They emphasized the importance of designing these resources inclusively rather than exclusively. Inclusive design prioritizes accessibility from the outset, creating products that naturally integrate the needs of all users without requiring additional attachments or modifications that could unintentionally isolate BVI individuals \cite{steinfeld2012universal}. StreetScape embodies this philosophy by addressing accessibility from two key perspectives. First, it empowers BVI children to develop spatial understanding through an engaging and playful experience. Second, it facilitates meaningful interaction and collaboration between sighted and BVI children, breaking down social stigmas and fostering a shared understanding. By enabling both groups to engage with the puzzle in an inclusive way, StreetScape promotes interdependence and social cohesion through collaborative play.

\subsection{Design Layout}
Research underscores the importance of making learning enjoyable to sustain interest and avoid perceptions of learning as a chore \cite{malone1987making}. StreetScape incorporates gamified elements and provides a positive emotional incentive for children to explore spatial reasoning while encouraging friendly competition and communication among users. The tactile puzzle is designed around a 5x5 grid representing a hypothetical street map. This grid features modular components, including four-way intersections, straight and curved roads, and decorative elements such as streetlamps, stoplights, and buildings. The modular design allows for flexibility and creativity, enabling users to build varied layouts and preventing rote memorization of a single map. One user creates a custom map base by arranging slotted tiles—composed of road and blank pieces—while the other attempts to match the pieces and corresponding decorative elements to the base. To facilitate ease of assembly and ensure precise alignment, magnets are embedded within the tiles, simplifying the connection between components without relying on exact 3D-printed measurements. This approach supports accessibility and usability by reducing the need for intricate manual adjustments. 
The gamified design mandates that other user assemble the tiles of the top layer based on tactile feedback alone, without visual reference to the map.
The competitive element, combined with the collaborative creation of the map, fosters communication, interdependence, and mutual understanding between sighted and BVI children, aligning with the broader goals of StreetScape.

\subsection{Tactile Features}
\begin{figure}[h]
  \centering
  \includegraphics[scale = 0.35]{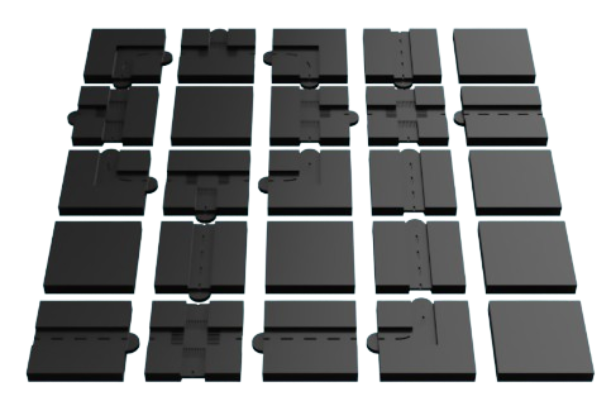}
  \caption{Freeform map made of slotted tiles that serves as the base for the tactile puzzle.}
  \Description{A 3D rendered image of road tiles serving as a puzzle base.}
   \label{fig:puzzlebase}
\end{figure}

\begin{figure}[h]
  \centering
  \includegraphics[scale = 0.7]{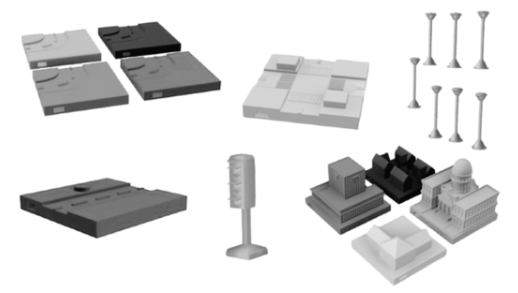}
  \caption{Overview of the different types of street tiles and decorative elements used in the system. The street tiles include straight roads, curved roads, and intersections, each featuring embedded magnet slots for easy attachment and modular reconfiguration. The tiles also incorporate different peg shapes—square pegs for stoplights and circular pegs for streetlamps—to facilitate tactile differentiation. Additional decorative tiles represent key urban structures, such as houses, a police station, a restaurant, and a school, along with freestanding streetlamps and stoplights to enhance engagement and spatial understanding.}
  \Description{A 3D rendered image of road tiles serving as a puzzle base.}
     \label{fig:decorative}
\end{figure}

To support intuitive and accessible interaction, StreetScape incorporates carefully designed tactile features that differentiate between base tiles and top tiles. As shown in Figure ~\ref{fig:puzzlebase}, the base tiles interlock using positive and negative joints, akin to a traditional puzzle, ensuring a secure and precise fit. This design facilitates the assembly of a wide range of map configurations, providing users with the flexibility to create new and engaging layouts, thereby sustaining interest over repeated gameplay. The top tiles, while lacking slotted joints, feature tactile road elements such as indents and curves that align seamlessly with the base pieces (see Figure~\ref{fig:decorative}). Roads are designed with distinct tactile cues: the road features are recessed into the tile, while raised, broken white lines run along the roads. These contrasting tactile elements ensure that users can distinguish road pieces from decorative engravings or other tile components purely through touch. 

For the decorative pieces, they’ve been exaggerated in order to highlight detail to the users. Stoplights have a square base, similar to certain stoplights found in real life, while the streetlights have circular bases (see Figure~\ref{fig:decorative}). This base differentiation is important since certain top tile pieces have differing square or circular pegs that the user needs to match to the corresponding piece, adding another challenging layer to the game. Decorative trees and buildings have flat bases that would correspond with any blank tiles placed in the base map.
\subsection{Development}
The development process followed an iterative design process, with each prototype addressing identified limitations and refining the tactile and interactive elements.

\subsubsection{Initial Conceptual Design}
\begin{figure}[h]
  \centering
  \includegraphics[scale = 0.7]{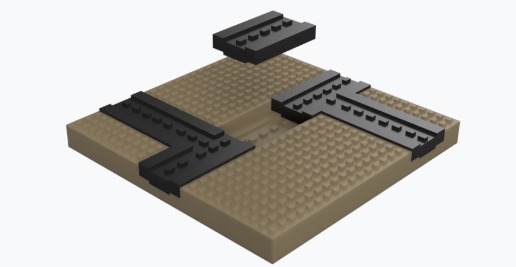}
  \caption{Details of the tile road tiles with slotted mechanisms}
  \Description{A 3D rendered image of road tiles}
  \label{fig:prototype1}
\end{figure}

The initial concept sought to emulate real-world map structures, focusing on roads and pathways to facilitate spatial learning for children. This early design featured fixed, preconstructed layouts with embedded pathways and prominent road features. Each tile incorporated slotted mechanisms and pegs to support decorative elements and customization, as illustrated in Figure \ref{fig:prototype1}. 
However, initial informal feedback from our collaborators highlighted that while the design effectively introduced structured spatial representations, it lacked opportunities for user-driven exploration and adaptability. This early version relied heavily on static configurations, limiting creative exploration and imaginative play. Children could memorize the fixed arrangement of tiles rather than actively engaging in dynamic assembly, reducing long-term engagement and educational value.

\subsubsection{Prototype 1: Sliding Slot Mechanism}
\begin{figure}[h]
  \centering
  \includegraphics[scale = 0.7]{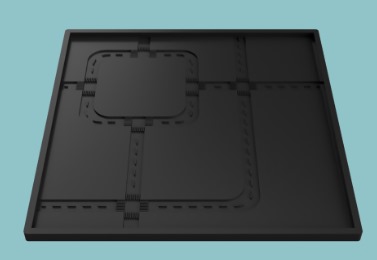} 
  \caption{Prototype 1 featuring the sliding slot mechanism, demonstrating how the tiles enhance tactile exploration.}
  \Description{A close-up view of the interlocking mechanism of the tactile puzzle tiles.}
  \label{fig:prototype1_1}
\end{figure}

The first physical prototype employed a sliding slot mechanism, which allowed tiles to interlock securely. Materials such as PLA, PETG, and carbon fiber-based filament were tested to evaluate durability, tactile clarity, and accessibility during assembly. While this design supported tactile exploration, as depicted in Figure \ref{fig:prototype1_1}, one of our collaborators provided critical feedback that helped shape the next iteration. They noted that the fixed puzzle-like structure constrained creativity, and the assembly process was not sufficiently flexible for BVI children to engage meaningfully with the map. Additionally, some children can have difficulty aligning the interlocking pieces due to the required fine motor coordination, presenting an accessibility barrier.

\subsubsection{Prototype 2: Modular and Magnetic Design}
\begin{figure}[h]
  \centering
  \includegraphics[scale=0.5]{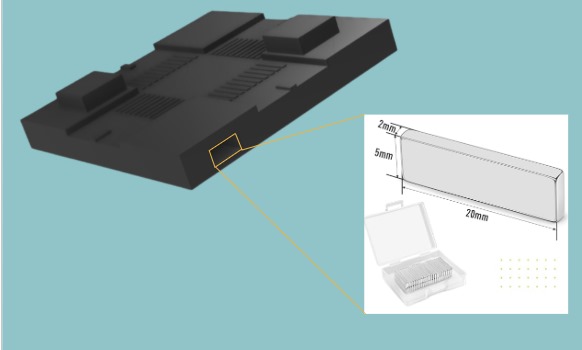} 
  \caption{Close-up of the embedded magnetic mechanisms used in Prototype 2 to facilitate easier and dynamic assembly of the puzzle pieces.}
  \Description{The image shows the detailed magnetic connections within the puzzle tiles, highlighting how they simplify the assembly process and enhance the flexibility of creating various map layouts.}
  \label{fig:prototype2}
\end{figure}

In response to these findings, the design transitioned to a modular system. The slotted mechanisms were replaced with embedded magnets, enabling easier and more dynamic assembly, illustrated in Figure \ref{fig:prototype2}. One player creates the base of the board using tile
pieces with positive and negative joints (see Figure \ref{fig:puzzlebase}). Second player has to magnetically match the top and
bottom tiles based on tactile feedback. 
This change allows users to create custom maps, fostering imaginative play and preventing rote memorization of fixed layouts. Additionally, the variety of decorative elements, such as miniature streetlamps, trees, and vehicles, was expanded to provide more opportunities for creative engagement and tactile exploration. The modularity here encourages collaborative play, as children could exchange and reposition tiles to create shared spatial narratives. 

\section{Discussion and Future Work}
StreetScape has the potential to offer an accessible and engaging platform for BVI children  to explore spatial relationships and collaborate with sighted peers. However, several limitations and opportunities for future work remain, which highlight the possibility for further refinement and exploration. Although the current design incorporates modular tiles, magnets for assembly, and distinct tactile markers, the addition of textured surfaces remains an unexplored area. Future iterations could include textures that simulate real-world materials, such as asphalt for roads, grass for parks, and bricks for buildings. These textures would provide a richer sensory experience, encouraging BVI children to interact with diverse surfaces and build a deeper understanding of their environment. Such enhancements would align with research emphasizing the importance of multisensory learning tools for children with disabilities. Additionally, while the design was informed by collaborators who are educators and individuals with blindness, a critical next step involves conducting formal studies with BVI children. These studies would provide invaluable insights into the usability, engagement, and educational impact of StreetScape. Specifically, evaluations could focus on: 
\begin{itemize}
\item The effectiveness of tactile features in enhancing spatial reasoning.
\item The impact of gamification on collaborative play between BVI and sighted children.
\item The suitability of the modular design for children with varying levels of motor skills.
\item The long-term engagement and replayability of StreetScape, particularly how children interact with the system over time and whether modular expansions sustain interest and learning benefits.
\end{itemize}

Such evaluations would also allow for iterative refinements based on direct user feedback, ensuring the puzzle meets the diverse needs of its target audience.

Future iterations could also consider adaptations for children with other disabilities, such as those with neurodiverse conditions or motor impairments. For instance, incorporating larger tiles or adjustable difficulty levels could make the puzzle more inclusive for a broader range of users. Additionally, tactile differentiation could be expanded to support children with both visual and cognitive impairments, ensuring that spatial learning remains intuitive across different ability levels. This aligns with the principles of universal design, which emphasize creating tools that are accessible to all individuals regardless of their abilities \cite{steinfeld2012universal}.

Another promising direction is the integration of auditory and haptic feedback to create a multi-sensory experience. For example, embedding small speakers or conductive materials into the tiles could enable audio-based feedback when specific roads, intersections, or landmarks are touched, reinforcing spatial learning through sound cues. Similarly, haptic vibrations could provide additional feedback when children correctly place pieces, guiding them through tactile confirmation. These multi-sensory elements would further enhance engagement, making StreetScape an adaptable and immersive tool.

A key strength of StreetScape lies in its ability to foster interdependence between sighted and BVI children. By designing for shared interaction rather than isolating accessibility features, the puzzle promotes mutual understanding and collaboration. Future studies could explore how this interdependence affects the social dynamics and perceptions of children participating in inclusive play. This research could provide broader insights into designing for collaborative accessibility in educational tools.

\section{Conclusion} 

This work presents StreetScape, a tactile street puzzle designed to promote spatial reasoning and collaborative play between BVI children and their sighted peers. By leveraging modular 3D-printed components, embedded magnets, and thoughtfully designed tactile features, the system provides an accessible and engaging platform for inclusive interaction. Throughout its development, StreetScape has been guided by principles of interdependence and inclusive design, addressing the need for tools that foster collaboration rather than isolation. The iterative design process enabled us to identify and refine critical elements, such as dynamic map configurations and intuitive tactile cues, ensuring the puzzle remains adaptable to various user needs. While the design was informed by collaborators with expertise in accessibility, formal user studies with BVI children remain an essential next step to evaluate and enhance its impact. Such evaluations will provide deeper insights into the puzzle's ability to support spatial learning, inclusivity, and social engagement. Through StreetScape, we contribute to the growing body of research on designing for accessibility, offering a practical and innovative approach to creating inclusive tools for BVI children. We hope this work inspires further exploration into the intersection of accessibility, gamification, and collaborative design, driving meaningful progress in human-computer interaction and inclusive education.

\bibliographystyle{ACM-Reference-Format}
\bibliography{sample-base}


\begin{thebibliography}{44}


\ifx \showCODEN    \undefined \def \showCODEN     #1{\unskip}     \fi
\ifx \showISBNx    \undefined \def \showISBNx     #1{\unskip}     \fi
\ifx \showISBNxiii \undefined \def \showISBNxiii  #1{\unskip}     \fi
\ifx \showISSN     \undefined \def \showISSN      #1{\unskip}     \fi
\ifx \showLCCN     \undefined \def \showLCCN      #1{\unskip}     \fi
\ifx \shownote     \undefined \def \shownote      #1{#1}          \fi
\ifx \showarticletitle \undefined \def \showarticletitle #1{#1}   \fi
\ifx \showURL      \undefined \def \showURL       {\relax}        \fi
\providecommand\bibfield[2]{#2}
\providecommand\bibinfo[2]{#2}
\providecommand\natexlab[1]{#1}
\providecommand\showeprint[2][]{arXiv:#2}

\bibitem[Batagoda et~al\mbox{.}(2023)]%
        {StepSure}
\bibfield{author}{\bibinfo{person}{Sadeepa Batagoda}, \bibinfo{person}{Pasindu Kolamunna}, \bibinfo{person}{H.~R. Sheheran}, \bibinfo{person}{D.~M. A. R.~G. Rajakaruna}, \bibinfo{person}{D.~S. Pandithage}, {and} \bibinfo{person}{Pipuni Wijesiri}.} \bibinfo{year}{2023}\natexlab{}.
\newblock \showarticletitle{StepSure: Smart Backpack for Blind Children Using IoT Device and Machine Learning}.
\newblock \bibinfo{journal}{\emph{International Journal of Research in Engineering, Science and Management}} \bibinfo{volume}{6}, \bibinfo{number}{11} (\bibinfo{date}{Nov.} \bibinfo{year}{2023}), \bibinfo{pages}{21–28}.
\newblock
\urldef\tempurl%
\url{https://journal.ijresm.com/index.php/ijresm/article/view/2849}
\showURL{%
\tempurl}


\bibitem[Bennett et~al\mbox{.}(2018)]%
        {Bennett2018}
\bibfield{author}{\bibinfo{person}{Cynthia~L. Bennett}, \bibinfo{person}{Erin Brady}, {and} \bibinfo{person}{Stacy~M. Branham}.} \bibinfo{year}{2018}\natexlab{}.
\newblock \showarticletitle{Interdependence as a Frame for Assistive Technology Research and Design}. In \bibinfo{booktitle}{\emph{Proceedings of the 20th International ACM SIGACCESS Conference on Computers and Accessibility}} (Galway, Ireland) \emph{(\bibinfo{series}{ASSETS '18})}. \bibinfo{publisher}{Association for Computing Machinery}, \bibinfo{address}{New York, NY, USA}, \bibinfo{pages}{161–173}.
\newblock
\showISBNx{9781450356503}
\href{https://doi.org/10.1145/3234695.3236348}{doi:\nolinkurl{10.1145/3234695.3236348}}


\bibitem[Brule et~al\mbox{.}(2016)]%
        {Brule2016}
\bibfield{author}{\bibinfo{person}{Emeline Brule}, \bibinfo{person}{Gilles Bailly}, \bibinfo{person}{Anke Brock}, \bibinfo{person}{Frederic Valentin}, \bibinfo{person}{Gr\'{e}goire Denis}, {and} \bibinfo{person}{Christophe Jouffrais}.} \bibinfo{year}{2016}\natexlab{}.
\newblock \showarticletitle{MapSense: Multi-Sensory Interactive Maps for Children Living with Visual Impairments}. In \bibinfo{booktitle}{\emph{Proceedings of the 2016 CHI Conference on Human Factors in Computing Systems}} (San Jose, California, USA) \emph{(\bibinfo{series}{CHI '16})}. \bibinfo{publisher}{Association for Computing Machinery}, \bibinfo{address}{New York, NY, USA}, \bibinfo{pages}{445–457}.
\newblock
\showISBNx{9781450333627}
\href{https://doi.org/10.1145/2858036.2858375}{doi:\nolinkurl{10.1145/2858036.2858375}}


\bibitem[Cahyono(2023)]%
        {Cahyono2023}
\bibfield{author}{\bibinfo{person}{Dony Cahyono}.} \bibinfo{year}{2023}\natexlab{}.
\newblock \showarticletitle{Gamification for Education: Using LexiPal to Foster Intrinsic and Extrinsic Learning Motivation of Students with Dyslexia}. In \bibinfo{booktitle}{\emph{Proceedings of the 14th International Conference on Education Technology and Computers}} (Barcelona, Spain) \emph{(\bibinfo{series}{ICETC '22})}. \bibinfo{publisher}{Association for Computing Machinery}, \bibinfo{address}{New York, NY, USA}, \bibinfo{pages}{32–38}.
\newblock
\showISBNx{9781450397766}
\href{https://doi.org/10.1145/3572549.3572555}{doi:\nolinkurl{10.1145/3572549.3572555}}


\bibitem[Cavazos~Quero et~al\mbox{.}(2019a)]%
        {Jido2019}
\bibfield{author}{\bibinfo{person}{Luis Cavazos~Quero}, \bibinfo{person}{Jorge Iranzo~Bartolom\'{e}}, \bibinfo{person}{Dongmyeong Lee}, \bibinfo{person}{Yerin Lee}, \bibinfo{person}{Sangwon Lee}, {and} \bibinfo{person}{Jundong Cho}.} \bibinfo{year}{2019}\natexlab{a}.
\newblock \showarticletitle{Jido: A Conversational Tactile Map for Blind People}. In \bibinfo{booktitle}{\emph{Proceedings of the 21st International ACM SIGACCESS Conference on Computers and Accessibility}} (Pittsburgh, PA, USA) \emph{(\bibinfo{series}{ASSETS '19})}. \bibinfo{publisher}{Association for Computing Machinery}, \bibinfo{address}{New York, NY, USA}, \bibinfo{pages}{682–684}.
\newblock
\showISBNx{9781450366762}
\href{https://doi.org/10.1145/3308561.3354600}{doi:\nolinkurl{10.1145/3308561.3354600}}


\bibitem[Cavazos~Quero et~al\mbox{.}(2019b)]%
        {jido}
\bibfield{author}{\bibinfo{person}{Luis Cavazos~Quero}, \bibinfo{person}{Jorge Iranzo~Bartolom\'{e}}, \bibinfo{person}{Dongmyeong Lee}, \bibinfo{person}{Yerin Lee}, \bibinfo{person}{Sangwon Lee}, {and} \bibinfo{person}{Jundong Cho}.} \bibinfo{year}{2019}\natexlab{b}.
\newblock \showarticletitle{Jido: A Conversational Tactile Map for Blind People}. In \bibinfo{booktitle}{\emph{Proceedings of the 21st International ACM SIGACCESS Conference on Computers and Accessibility}} (Pittsburgh, PA, USA) \emph{(\bibinfo{series}{ASSETS '19})}. \bibinfo{publisher}{Association for Computing Machinery}, \bibinfo{address}{New York, NY, USA}, \bibinfo{pages}{682–684}.
\newblock
\showISBNx{9781450366762}
\href{https://doi.org/10.1145/3308561.3354600}{doi:\nolinkurl{10.1145/3308561.3354600}}


\bibitem[Coughlan et~al\mbox{.}(2022)]%
        {Coughlan2022}
\bibfield{author}{\bibinfo{person}{James~M. Coughlan}, \bibinfo{person}{Brandon Biggs}, {and} \bibinfo{person}{Huiying Shen}.} \bibinfo{year}{2022}\natexlab{}.
\newblock \showarticletitle{Non-visual Access to an Interactive 3D Map}. In \bibinfo{booktitle}{\emph{Computers Helping People with Special Needs: 18th International Conference, ICCHP-AAATE 2022, Lecco, Italy, July 11–15, 2022, Proceedings, Part I}} (Milan, Italy). \bibinfo{publisher}{Springer-Verlag}, \bibinfo{address}{Berlin, Heidelberg}, \bibinfo{pages}{253–260}.
\newblock
\showISBNx{978-3-031-08647-2}
\href{https://doi.org/10.1007/978-3-031-08648-9_29}{doi:\nolinkurl{10.1007/978-3-031-08648-9_29}}


\bibitem[Crawford et~al\mbox{.}(2024)]%
        {Crawford2024}
\bibfield{author}{\bibinfo{person}{Kirk~Andrew Crawford}, \bibinfo{person}{Jennifer Posada}, \bibinfo{person}{Yetunde~Esther Okueso}, \bibinfo{person}{Erin Higgins}, \bibinfo{person}{Laura Lachin}, {and} \bibinfo{person}{Foad Hamidi}.} \bibinfo{year}{2024}\natexlab{}.
\newblock \showarticletitle{Co-designing a 3D-Printed Tactile Campus Map With Blind and Low-Vision University Students}. In \bibinfo{booktitle}{\emph{Proceedings of the 26th International ACM SIGACCESS Conference on Computers and Accessibility}} (St. John's, NL, Canada) \emph{(\bibinfo{series}{ASSETS '24})}. \bibinfo{publisher}{Association for Computing Machinery}, \bibinfo{address}{New York, NY, USA}, Article \bibinfo{articleno}{77}, \bibinfo{numpages}{6}~pages.
\newblock
\showISBNx{9798400706776}
\href{https://doi.org/10.1145/3663548.3688537}{doi:\nolinkurl{10.1145/3663548.3688537}}


\bibitem[Deterding et~al\mbox{.}(2011)]%
        {gamification}
\bibfield{author}{\bibinfo{person}{Sebastian Deterding}, \bibinfo{person}{Dan Dixon}, \bibinfo{person}{Rilla Khaled}, {and} \bibinfo{person}{Lennart Nacke}.} \bibinfo{year}{2011}\natexlab{}.
\newblock \showarticletitle{From game design elements to gamefulness: defining "gamification"}. In \bibinfo{booktitle}{\emph{Proceedings of the 15th International Academic MindTrek Conference: Envisioning Future Media Environments}} (Tampere, Finland) \emph{(\bibinfo{series}{MindTrek '11})}. \bibinfo{publisher}{Association for Computing Machinery}, \bibinfo{address}{New York, NY, USA}, \bibinfo{pages}{9–15}.
\newblock
\showISBNx{9781450308168}
\href{https://doi.org/10.1145/2181037.2181040}{doi:\nolinkurl{10.1145/2181037.2181040}}


\bibitem[Espinosa et~al\mbox{.}(1998)]%
        {cogprints1509}
\bibfield{author}{\bibinfo{person}{Maria~Angeles Espinosa}, \bibinfo{person}{Simon Ungar}, \bibinfo{person}{Esperanza Ochaita}, \bibinfo{person}{Mark Blades}, {and} \bibinfo{person}{Christopher Spencer}.} \bibinfo{year}{1998}\natexlab{}.
\newblock \bibinfo{title}{Comparing methods for introducing blind and visually impaired people to unfamiliar urban environments}.
\newblock \bibinfo{numpages}{277--287}~pages.
\newblock
\urldef\tempurl%
\url{http://cogprints.org/1509/}
\showURL{%
\tempurl}


\bibitem[Fielder et~al\mbox{.}(1993)]%
        {Fielder1993}
\bibfield{author}{\bibinfo{person}{A. Fielder}, \bibinfo{person}{A. Best}, {and} \bibinfo{person}{M. Bax}.} \bibinfo{year}{1993}\natexlab{}.
\newblock \bibinfo{booktitle}{\emph{The Management of Visual Impairment in Childhood}}.
\newblock \bibinfo{publisher}{Mac Keith Press}.
\newblock


\bibitem[Freeman et~al\mbox{.}(2017)]%
        {Freeman2017}
\bibfield{author}{\bibinfo{person}{Euan Freeman}, \bibinfo{person}{Graham Wilson}, \bibinfo{person}{Stephen Brewster}, \bibinfo{person}{Gabriel Baud-Bovy}, \bibinfo{person}{Charlotte Magnusson}, {and} \bibinfo{person}{Hector Caltenco}.} \bibinfo{year}{2017}\natexlab{}.
\newblock \showarticletitle{Audible Beacons and Wearables in Schools: Helping Young Visually Impaired Children Play and Move Independently}. In \bibinfo{booktitle}{\emph{Proceedings of the 2017 CHI Conference on Human Factors in Computing Systems}} (Denver, Colorado, USA) \emph{(\bibinfo{series}{CHI '17})}. \bibinfo{publisher}{Association for Computing Machinery}, \bibinfo{address}{New York, NY, USA}, \bibinfo{pages}{4146–4157}.
\newblock
\showISBNx{9781450346559}
\href{https://doi.org/10.1145/3025453.3025518}{doi:\nolinkurl{10.1145/3025453.3025518}}


\bibitem[Gadiraju et~al\mbox{.}(2024)]%
        {Gadiraju2024}
\bibfield{author}{\bibinfo{person}{Vinitha Gadiraju}, \bibinfo{person}{Lucia Jayne}, {and} \bibinfo{person}{Shaun~K. Kane}.} \bibinfo{year}{2024}\natexlab{}.
\newblock \showarticletitle{"It's an independent living skill, but covered with fun!": Prompting At-Home Skill Development for Children with Vision Impairment}. In \bibinfo{booktitle}{\emph{Proceedings of the 26th International ACM SIGACCESS Conference on Computers and Accessibility}} (St. John's, NL, Canada) \emph{(\bibinfo{series}{ASSETS '24})}. \bibinfo{publisher}{Association for Computing Machinery}, \bibinfo{address}{New York, NY, USA}, Article \bibinfo{articleno}{27}, \bibinfo{numpages}{14}~pages.
\newblock
\showISBNx{9798400706776}
\href{https://doi.org/10.1145/3663548.3675626}{doi:\nolinkurl{10.1145/3663548.3675626}}


\bibitem[Gon\c{c}alves et~al\mbox{.}(2023)]%
        {Manuel2023}
\bibfield{author}{\bibinfo{person}{David Gon\c{c}alves}, \bibinfo{person}{Manuel Pi\c{c}arra}, \bibinfo{person}{Pedro Pais}, \bibinfo{person}{Jo\~{a}o Guerreiro}, {and} \bibinfo{person}{Andr\'{e} Rodrigues}.} \bibinfo{year}{2023}\natexlab{}.
\newblock \showarticletitle{"My Zelda Cane": Strategies Used by Blind Players to Play Visual-Centric Digital Games}. In \bibinfo{booktitle}{\emph{Proceedings of the 2023 CHI Conference on Human Factors in Computing Systems}} (Hamburg, Germany) \emph{(\bibinfo{series}{CHI '23})}. \bibinfo{publisher}{Association for Computing Machinery}, \bibinfo{address}{New York, NY, USA}, Article \bibinfo{articleno}{289}, \bibinfo{numpages}{15}~pages.
\newblock
\showISBNx{9781450394215}
\href{https://doi.org/10.1145/3544548.3580702}{doi:\nolinkurl{10.1145/3544548.3580702}}


\bibitem[G\"{o}tzelmann(2016)]%
        {LucentMaps}
\bibfield{author}{\bibinfo{person}{Timo G\"{o}tzelmann}.} \bibinfo{year}{2016}\natexlab{}.
\newblock \showarticletitle{LucentMaps: 3D Printed Audiovisual Tactile Maps for Blind and Visually Impaired People}. In \bibinfo{booktitle}{\emph{Proceedings of the 18th International ACM SIGACCESS Conference on Computers and Accessibility}} (Reno, Nevada, USA) \emph{(\bibinfo{series}{ASSETS '16})}. \bibinfo{publisher}{Association for Computing Machinery}, \bibinfo{address}{New York, NY, USA}, \bibinfo{pages}{81–90}.
\newblock
\showISBNx{9781450341240}
\href{https://doi.org/10.1145/2982142.2982163}{doi:\nolinkurl{10.1145/2982142.2982163}}


\bibitem[Guerreiro et~al\mbox{.}(2019)]%
        {Guerreiro2019}
\bibfield{author}{\bibinfo{person}{Jo\~{a}o Guerreiro}, \bibinfo{person}{Dragan Ahmetovic}, \bibinfo{person}{Daisuke Sato}, \bibinfo{person}{Kris Kitani}, {and} \bibinfo{person}{Chieko Asakawa}.} \bibinfo{year}{2019}\natexlab{}.
\newblock \showarticletitle{Airport Accessibility and Navigation Assistance for People with Visual Impairments}. In \bibinfo{booktitle}{\emph{Proceedings of the 2019 CHI Conference on Human Factors in Computing Systems}} (Glasgow, Scotland Uk) \emph{(\bibinfo{series}{CHI '19})}. \bibinfo{publisher}{Association for Computing Machinery}, \bibinfo{address}{New York, NY, USA}, \bibinfo{pages}{1–14}.
\newblock
\showISBNx{9781450359702}
\href{https://doi.org/10.1145/3290605.3300246}{doi:\nolinkurl{10.1145/3290605.3300246}}


\bibitem[Gupta et~al\mbox{.}(2020)]%
        {Gupta2020}
\bibfield{author}{\bibinfo{person}{Maya Gupta}, \bibinfo{person}{Ali Abdolrahmani}, \bibinfo{person}{Emory Edwards}, \bibinfo{person}{Mayra Cortez}, \bibinfo{person}{Andrew Tumang}, \bibinfo{person}{Yasmin Majali}, \bibinfo{person}{Marc Lazaga}, \bibinfo{person}{Samhitha Tarra}, \bibinfo{person}{Prasad Patil}, \bibinfo{person}{Ravi Kuber}, {and} \bibinfo{person}{Stacy~M. Branham}.} \bibinfo{year}{2020}\natexlab{}.
\newblock \showarticletitle{Towards More Universal Wayfinding Technologies: Navigation Preferences Across Disabilities}. In \bibinfo{booktitle}{\emph{Proceedings of the 2020 CHI Conference on Human Factors in Computing Systems}} (Honolulu, HI, USA) \emph{(\bibinfo{series}{CHI '20})}. \bibinfo{publisher}{Association for Computing Machinery}, \bibinfo{address}{New York, NY, USA}, \bibinfo{pages}{1–13}.
\newblock
\showISBNx{9781450367080}
\href{https://doi.org/10.1145/3313831.3376581}{doi:\nolinkurl{10.1145/3313831.3376581}}


\bibitem[Hofmann et~al\mbox{.}(2022)]%
        {Hofmann2022}
\bibfield{author}{\bibinfo{person}{Megan Hofmann}, \bibinfo{person}{Kelly Mack}, \bibinfo{person}{Jessica Birchfield}, \bibinfo{person}{Jerry Cao}, \bibinfo{person}{Autumn~G Hughes}, \bibinfo{person}{Shriya Kurpad}, \bibinfo{person}{Kathryn~J Lum}, \bibinfo{person}{Emily Warnock}, \bibinfo{person}{Anat Caspi}, \bibinfo{person}{Scott~E Hudson}, {and} \bibinfo{person}{Jennifer Mankoff}.} \bibinfo{year}{2022}\natexlab{}.
\newblock \showarticletitle{Maptimizer: Using Optimization to Tailor Tactile Maps to Users Needs}. In \bibinfo{booktitle}{\emph{Proceedings of the 2022 CHI Conference on Human Factors in Computing Systems}} (New Orleans, LA, USA) \emph{(\bibinfo{series}{CHI '22})}. \bibinfo{publisher}{Association for Computing Machinery}, \bibinfo{address}{New York, NY, USA}, Article \bibinfo{articleno}{592}, \bibinfo{numpages}{15}~pages.
\newblock
\showISBNx{9781450391573}
\href{https://doi.org/10.1145/3491102.3517436}{doi:\nolinkurl{10.1145/3491102.3517436}}


\bibitem[Holbrook and Koenig(2000)]%
        {holbrook2000foundations}
\bibfield{editor}{\bibinfo{person}{M.~Cay Holbrook} {and} \bibinfo{person}{Alan~J. Koenig}} (Eds.). \bibinfo{year}{2000}\natexlab{}.
\newblock \bibinfo{booktitle}{\emph{Foundations of Education: Instructional strategies for teaching children and youths with visual impairments}}. \bibinfo{series}{Foundations series}, Vol.~\bibinfo{volume}{2}.
\newblock \bibinfo{publisher}{American Foundation for the Blind}. 335 pages.
\newblock
\showISBNx{0891283390, 9780891283393}


\bibitem[Hussein et~al\mbox{.}(2023)]%
        {hussein2023gamification}
\bibfield{author}{\bibinfo{person}{E. Hussein}, \bibinfo{person}{A. Kan’an}, \bibinfo{person}{A. Rasheed}, \bibinfo{person}{Y. Alrashed}, \bibinfo{person}{M. Jdaitawi}, \bibinfo{person}{A. Abas}, \bibinfo{person}{S. Mabrouk}, {and} \bibinfo{person}{M. Abdelmoneim}.} \bibinfo{year}{2023}\natexlab{}.
\newblock \showarticletitle{Exploring the impact of gamification on skill development in special education: A systematic review}.
\newblock \bibinfo{journal}{\emph{Contemporary Educational Technology}} \bibinfo{volume}{15}, \bibinfo{number}{3} (\bibinfo{year}{2023}), \bibinfo{pages}{ep443}.
\newblock
\href{https://doi.org/10.30935/cedtech/13335}{doi:\nolinkurl{10.30935/cedtech/13335}}


\bibitem[Jacobson(1992)]%
        {Jacobson1992}
\bibfield{author}{\bibinfo{person}{Dan Jacobson}.} \bibinfo{year}{1992}\natexlab{}.
\newblock \showarticletitle{Spatial Cognition Through Tactile Mapping}.
\newblock \bibinfo{journal}{\emph{Swansea Geographer}}  \bibinfo{volume}{29} (\bibinfo{date}{01} \bibinfo{year}{1992}).
\newblock


\bibitem[Kaarlela(1959)]%
        {Ruth1959}
\bibfield{author}{\bibinfo{person}{Ruth Kaarlela}.} \bibinfo{year}{1959}\natexlab{}.
\newblock \showarticletitle{The Role of the Family in Developing Independence in the Blind Child}.
\newblock \bibinfo{journal}{\emph{Journal of Visual Impairment \& Blindness}} \bibinfo{volume}{53}, \bibinfo{number}{7} (\bibinfo{year}{1959}), \bibinfo{pages}{245--248}.
\newblock
\href{https://doi.org/10.1177/0145482X5905300702}{doi:\nolinkurl{10.1177/0145482X5905300702}}
\showeprint{https://doi.org/10.1177/0145482X5905300702}


\bibitem[Kayukawa et~al\mbox{.}(2023)]%
        {Kayukawa2023}
\bibfield{author}{\bibinfo{person}{Seita Kayukawa}, \bibinfo{person}{Daisuke Sato}, \bibinfo{person}{Masayuki Murata}, \bibinfo{person}{Tatsuya Ishihara}, \bibinfo{person}{Hironobu Takagi}, \bibinfo{person}{Shigeo Morishima}, {and} \bibinfo{person}{Chieko Asakawa}.} \bibinfo{year}{2023}\natexlab{}.
\newblock \showarticletitle{Enhancing Blind Visitor’s Autonomy in a Science Museum Using an Autonomous Navigation Robot}. In \bibinfo{booktitle}{\emph{Proceedings of the 2023 CHI Conference on Human Factors in Computing Systems}} (Hamburg, Germany) \emph{(\bibinfo{series}{CHI '23})}. \bibinfo{publisher}{Association for Computing Machinery}, \bibinfo{address}{New York, NY, USA}, Article \bibinfo{articleno}{541}, \bibinfo{numpages}{14}~pages.
\newblock
\showISBNx{9781450394215}
\href{https://doi.org/10.1145/3544548.3581220}{doi:\nolinkurl{10.1145/3544548.3581220}}


\bibitem[Kiryakova et~al\mbox{.}(2014)]%
        {Kiryakova}
\bibfield{author}{\bibinfo{person}{Gabriela Kiryakova}, \bibinfo{person}{Nadezhda Angelova}, {and} \bibinfo{person}{Lina Yordanova}.} \bibinfo{year}{2014}\natexlab{}.
\newblock \showarticletitle{GAMIFICATION IN EDUCATION}.
\newblock
\urldef\tempurl%
\url{https://www.researchgate.net/publication/320234774}
\showURL{%
\tempurl}


\bibitem[Kubota et~al\mbox{.}(2024)]%
        {Kuribayashi2024}
\bibfield{author}{\bibinfo{person}{Masaya Kubota}, \bibinfo{person}{Masaki Kuribayashi}, \bibinfo{person}{Seita Kayukawa}, \bibinfo{person}{Hironobu Takagi}, \bibinfo{person}{Chieko Asakawa}, {and} \bibinfo{person}{Shigeo Morishima}.} \bibinfo{year}{2024}\natexlab{}.
\newblock \showarticletitle{Snap and Nav: Smartphone-based Indoor Navigation System For Blind People via Floor Map Analysis and Intersection Detection}.
\newblock \bibinfo{journal}{\emph{Proc. ACM Hum.-Comput. Interact.}} \bibinfo{volume}{8}, \bibinfo{number}{MHCI}, Article \bibinfo{articleno}{275} (\bibinfo{date}{Sept.} \bibinfo{year}{2024}), \bibinfo{numpages}{22}~pages.
\newblock
\href{https://doi.org/10.1145/3676522}{doi:\nolinkurl{10.1145/3676522}}


\bibitem[Kuribayashi et~al\mbox{.}(2021)]%
        {Kuribayashi2021}
\bibfield{author}{\bibinfo{person}{Masaki Kuribayashi}, \bibinfo{person}{Seita Kayukawa}, \bibinfo{person}{Hironobu Takagi}, \bibinfo{person}{Chieko Asakawa}, {and} \bibinfo{person}{Shigeo Morishima}.} \bibinfo{year}{2021}\natexlab{}.
\newblock \showarticletitle{LineChaser: A Smartphone-Based Navigation System for Blind People to Stand in Lines}. In \bibinfo{booktitle}{\emph{Proceedings of the 2021 CHI Conference on Human Factors in Computing Systems}} (Yokohama, Japan) \emph{(\bibinfo{series}{CHI '21})}. \bibinfo{publisher}{Association for Computing Machinery}, \bibinfo{address}{New York, NY, USA}, Article \bibinfo{articleno}{33}, \bibinfo{numpages}{13}~pages.
\newblock
\showISBNx{9781450380966}
\href{https://doi.org/10.1145/3411764.3445451}{doi:\nolinkurl{10.1145/3411764.3445451}}


\bibitem[Landau et~al\mbox{.}(1984)]%
        {landau1984spatial}
\bibfield{author}{\bibinfo{person}{Barbara Landau}, \bibinfo{person}{Elizabeth Spelke}, {and} \bibinfo{person}{Henry Gleitman}.} \bibinfo{year}{1984}\natexlab{}.
\newblock \showarticletitle{Spatial knowledge in a young blind child}.
\newblock \bibinfo{journal}{\emph{Cognition}} \bibinfo{volume}{16}, \bibinfo{number}{3} (\bibinfo{year}{1984}), \bibinfo{pages}{225--260}.
\newblock
\href{https://doi.org/10.1016/0010-0277(84)90005-0}{doi:\nolinkurl{10.1016/0010-0277(84)90005-0}}


\bibitem[Mahmoudi et~al\mbox{.}(2024)]%
        {mahmoudi2024gamification}
\bibfield{author}{\bibinfo{person}{E. Mahmoudi}, \bibinfo{person}{P.~Yejong Yoo}, \bibinfo{person}{A. Chandra}, \bibinfo{person}{R. Cardoso}, \bibinfo{person}{C. Denner Dos~Santos}, \bibinfo{person}{A. Majnemer}, {and} \bibinfo{person}{K. Shikako}.} \bibinfo{year}{2024}\natexlab{}.
\newblock \showarticletitle{Gamification in Mobile Apps for Children With Disabilities: Scoping Review}.
\newblock \bibinfo{journal}{\emph{JMIR Serious Games}}  \bibinfo{volume}{12} (\bibinfo{date}{Sep 6} \bibinfo{year}{2024}), \bibinfo{pages}{e49029}.
\newblock
\href{https://doi.org/10.2196/49029}{doi:\nolinkurl{10.2196/49029}}


\bibitem[Malone and Lepper(1987)]%
        {malone1987making}
\bibfield{author}{\bibinfo{person}{Thomas~W. Malone} {and} \bibinfo{person}{Mark~R. Lepper}.} \bibinfo{year}{1987}\natexlab{}.
\newblock \showarticletitle{Making Learning Fun: A Taxonomy of Intrinsic Motivations for Learning}.
\newblock In \bibinfo{booktitle}{\emph{Aptitude, Learning, and Instruction} (\bibinfo{edition}{1st} ed.)}. \bibinfo{publisher}{Routledge}, \bibinfo{pages}{32}.
\newblock
\showISBNx{9781003163244}


\bibitem[Mikulowski(2019)]%
        {Mikulowski2019}
\bibfield{author}{\bibinfo{person}{Dariusz Mikulowski}.} \bibinfo{year}{2019}\natexlab{}.
\newblock \showarticletitle{An Approach to Teaching Blind Children of Geographic Topics Through Applying a Combined Multimodal User Interfaces}. In \bibinfo{booktitle}{\emph{Smart Industry {\&} Smart Education}}, \bibfield{editor}{\bibinfo{person}{Michael~E. Auer} {and} \bibinfo{person}{Reinhard Langmann}} (Eds.). \bibinfo{publisher}{Springer International Publishing}, \bibinfo{address}{Cham}, \bibinfo{pages}{435--442}.
\newblock
\showISBNx{978-3-319-95678-7}


\bibitem[Pyle et~al\mbox{.}(2017)]%
        {pyle2017play}
\bibfield{author}{\bibinfo{person}{Angela Pyle}, \bibinfo{person}{Christopher DeLuca}, {and} \bibinfo{person}{Erika Danniels}.} \bibinfo{year}{2017}\natexlab{}.
\newblock \showarticletitle{A scoping review of research on play-based pedagogies in kindergarten education}.
\newblock \bibinfo{journal}{\emph{Review of Education}} \bibinfo{volume}{5}, \bibinfo{number}{3} (\bibinfo{year}{2017}), \bibinfo{pages}{311--351}.
\newblock
\href{https://doi.org/10.1002/rev3.3097}{doi:\nolinkurl{10.1002/rev3.3097}}


\bibitem[Rocha et~al\mbox{.}(2021)]%
        {Rocha2021}
\bibfield{author}{\bibinfo{person}{Filipa Rocha}, \bibinfo{person}{Ana~Cristina Pires}, \bibinfo{person}{Isabel Neto}, \bibinfo{person}{Hugo Nicolau}, {and} \bibinfo{person}{Tiago Guerreiro}.} \bibinfo{year}{2021}\natexlab{}.
\newblock \showarticletitle{Accembly at Home: Accessible Spatial Programming for Children with Visual Impairments and their Families}. In \bibinfo{booktitle}{\emph{Proceedings of the 20th Annual ACM Interaction Design and Children Conference}} (Athens, Greece) \emph{(\bibinfo{series}{IDC '21})}. \bibinfo{publisher}{Association for Computing Machinery}, \bibinfo{address}{New York, NY, USA}, \bibinfo{pages}{100–111}.
\newblock
\showISBNx{9781450384520}
\href{https://doi.org/10.1145/3459990.3460699}{doi:\nolinkurl{10.1145/3459990.3460699}}


\bibitem[S\'{a}nchez et~al\mbox{.}(2010)]%
        {Jaime2010}
\bibfield{author}{\bibinfo{person}{Jaime S\'{a}nchez}, \bibinfo{person}{Mauricio Saenz}, {and} \bibinfo{person}{Jose~Miguel Garrido}.} \bibinfo{year}{2010}\natexlab{}.
\newblock \showarticletitle{Usability of a Multimodal Video Game to Improve Navigation Skills for Blind Children}.
\newblock \bibinfo{journal}{\emph{ACM Trans. Access. Comput.}} \bibinfo{volume}{3}, \bibinfo{number}{2}, Article \bibinfo{articleno}{7} (\bibinfo{date}{Nov.} \bibinfo{year}{2010}), \bibinfo{numpages}{29}~pages.
\newblock
\showISSN{1936-7228}
\href{https://doi.org/10.1145/1857920.1857924}{doi:\nolinkurl{10.1145/1857920.1857924}}


\bibitem[Shen et~al\mbox{.}(2013)]%
        {CamIO2013}
\bibfield{author}{\bibinfo{person}{Huiying Shen}, \bibinfo{person}{Owen Edwards}, \bibinfo{person}{Joshua Miele}, {and} \bibinfo{person}{James~M. Coughlan}.} \bibinfo{year}{2013}\natexlab{}.
\newblock \showarticletitle{CamIO: a 3D computer vision system enabling audio/haptic interaction with physical objects by blind users}. In \bibinfo{booktitle}{\emph{Proceedings of the 15th International ACM SIGACCESS Conference on Computers and Accessibility}} (Bellevue, Washington) \emph{(\bibinfo{series}{ASSETS '13})}. \bibinfo{publisher}{Association for Computing Machinery}, \bibinfo{address}{New York, NY, USA}, Article \bibinfo{articleno}{41}, \bibinfo{numpages}{2}~pages.
\newblock
\showISBNx{9781450324052}
\href{https://doi.org/10.1145/2513383.2513423}{doi:\nolinkurl{10.1145/2513383.2513423}}


\bibitem[Sobel et~al\mbox{.}(2016)]%
        {Sobel2016}
\bibfield{author}{\bibinfo{person}{Kiley Sobel}, \bibinfo{person}{Kyle Rector}, \bibinfo{person}{Susan Evans}, {and} \bibinfo{person}{Julie~A. Kientz}.} \bibinfo{year}{2016}\natexlab{}.
\newblock \showarticletitle{Incloodle: Evaluating an Interactive Application for Young Children with Mixed Abilities}. In \bibinfo{booktitle}{\emph{Proceedings of the 2016 CHI Conference on Human Factors in Computing Systems}} (San Jose, California, USA) \emph{(\bibinfo{series}{CHI '16})}. \bibinfo{publisher}{Association for Computing Machinery}, \bibinfo{address}{New York, NY, USA}, \bibinfo{pages}{165–176}.
\newblock
\showISBNx{9781450333627}
\href{https://doi.org/10.1145/2858036.2858114}{doi:\nolinkurl{10.1145/2858036.2858114}}


\bibitem[Steinfeld and Maisel(2012)]%
        {steinfeld2012universal}
\bibfield{author}{\bibinfo{person}{Edward Steinfeld} {and} \bibinfo{person}{Jordana Maisel}.} \bibinfo{year}{2012}\natexlab{}.
\newblock \bibinfo{booktitle}{\emph{Universal Design: Creating Inclusive Environments}}.
\newblock \bibinfo{publisher}{Wiley}, \bibinfo{address}{Hoboken, NJ}.
\newblock


\bibitem[Swallow and Huebner(1987)]%
        {swallow1987thrive}
\bibfield{author}{\bibinfo{person}{Rose-Marie Swallow} {and} \bibinfo{person}{Kathleen~Mary Huebner}.} \bibinfo{year}{1987}\natexlab{}.
\newblock \bibinfo{booktitle}{\emph{How to thrive, not just survive: A guide to developing independent life skills for blind and visually impaired children and youths}}.
\newblock \bibinfo{publisher}{American Foundation for the Blind}.
\newblock


\bibitem[Sánchez and Flores(2005)]%
        {AudioMath}
\bibfield{author}{\bibinfo{person}{Jaime Sánchez} {and} \bibinfo{person}{Hector Flores}.} \bibinfo{year}{2005}\natexlab{}.
\newblock \showarticletitle{AudioMath: Blind children learning mathematics through audio}.
\newblock \bibinfo{journal}{\emph{International Journal on Disability and Human Development}} \bibinfo{volume}{4}, \bibinfo{number}{4} (\bibinfo{year}{2005}), \bibinfo{pages}{311--316}.
\newblock
\href{https://doi.org/doi:10.1515/IJDHD.2005.4.4.311}{doi:\nolinkurl{doi:10.1515/IJDHD.2005.4.4.311}}


\bibitem[Taylor et~al\mbox{.}(2016)]%
        {Taylor2016}
\bibfield{author}{\bibinfo{person}{Brandon Taylor}, \bibinfo{person}{Anind Dey}, \bibinfo{person}{Dan Siewiorek}, {and} \bibinfo{person}{Asim Smailagic}.} \bibinfo{year}{2016}\natexlab{}.
\newblock \showarticletitle{Customizable 3D Printed Tactile Maps as Interactive Overlays}. In \bibinfo{booktitle}{\emph{Proceedings of the 18th International ACM SIGACCESS Conference on Computers and Accessibility}} (Reno, Nevada, USA) \emph{(\bibinfo{series}{ASSETS '16})}. \bibinfo{publisher}{Association for Computing Machinery}, \bibinfo{address}{New York, NY, USA}, \bibinfo{pages}{71–79}.
\newblock
\showISBNx{9781450341240}
\href{https://doi.org/10.1145/2982142.2982167}{doi:\nolinkurl{10.1145/2982142.2982167}}


\bibitem[Wei et~al\mbox{.}(2022)]%
        {Wei2022}
\bibfield{author}{\bibinfo{person}{Linchao Wei}, \bibinfo{person}{Lingling Jin}, \bibinfo{person}{Ruining Gong}, \bibinfo{person}{Yaojun Yang}, {and} \bibinfo{person}{Xiaochen Zhang}.} \bibinfo{year}{2022}\natexlab{}.
\newblock \showarticletitle{Design of Audio-Augmented-Reality-Based O\&M Orientation Training for Visually Impaired Children}.
\newblock \bibinfo{journal}{\emph{Sensors}} \bibinfo{volume}{22}, \bibinfo{number}{23} (\bibinfo{year}{2022}).
\newblock
\showISSN{1424-8220}
\href{https://doi.org/10.3390/s22239487}{doi:\nolinkurl{10.3390/s22239487}}


\bibitem[Williams et~al\mbox{.}(2013)]%
        {Williams2013}
\bibfield{author}{\bibinfo{person}{Michele~A. Williams}, \bibinfo{person}{Amy Hurst}, {and} \bibinfo{person}{Shaun~K. Kane}.} \bibinfo{year}{2013}\natexlab{}.
\newblock \showarticletitle{"Pray before you step out": describing personal and situational blind navigation behaviors}. In \bibinfo{booktitle}{\emph{Proceedings of the 15th International ACM SIGACCESS Conference on Computers and Accessibility}} (Bellevue, Washington) \emph{(\bibinfo{series}{ASSETS '13})}. \bibinfo{publisher}{Association for Computing Machinery}, \bibinfo{address}{New York, NY, USA}, Article \bibinfo{articleno}{28}, \bibinfo{numpages}{8}~pages.
\newblock
\showISBNx{9781450324052}
\href{https://doi.org/10.1145/2513383.2513449}{doi:\nolinkurl{10.1145/2513383.2513449}}


\bibitem[{World Health Organization}(2015)]%
        {WHO2015}
\bibfield{author}{\bibinfo{person}{{World Health Organization}}.} \bibinfo{year}{2015}\natexlab{}.
\newblock \bibinfo{title}{Visual disturbances and blindness (H53-H54)}.
\newblock \bibinfo{howpublished}{\url{https://www.who.int}}.
\newblock
\newblock
\shownote{[Online; accessed 10-January-2025]}.


\bibitem[Yogman et~al\mbox{.}(2018)]%
        {yogman2018power}
\bibfield{author}{\bibinfo{person}{Michael Yogman}, \bibinfo{person}{Andrew Garner}, \bibinfo{person}{Jeffrey Hutchinson}, \bibinfo{person}{Kathy Hirsh-Pasek}, \bibinfo{person}{Roberta~Michnick Golinkoff}, \bibinfo{person}{Rebecca Baum}, \bibinfo{person}{Thresia Gambon}, \bibinfo{person}{Arthur Lavin}, \bibinfo{person}{Gerri Mattson}, \bibinfo{person}{Lawrence Wissow}, {et~al\mbox{.}}} \bibinfo{year}{2018}\natexlab{}.
\newblock \showarticletitle{The power of play: A pediatric role in enhancing development in young children}.
\newblock \bibinfo{journal}{\emph{Pediatrics}} \bibinfo{volume}{142}, \bibinfo{number}{3} (\bibinfo{year}{2018}).
\newblock


\bibitem[Zhao et~al\mbox{.}(2020)]%
        {Kupferstein2020}
\bibfield{author}{\bibinfo{person}{Yuhang Zhao}, \bibinfo{person}{Elizabeth Kupferstein}, \bibinfo{person}{Hathaitorn Rojnirun}, \bibinfo{person}{Leah Findlater}, {and} \bibinfo{person}{Shiri Azenkot}.} \bibinfo{year}{2020}\natexlab{}.
\newblock \showarticletitle{The Effectiveness of Visual and Audio Wayfinding Guidance on Smartglasses for People with Low Vision}. In \bibinfo{booktitle}{\emph{Proceedings of the 2020 CHI Conference on Human Factors in Computing Systems}} (Honolulu, HI, USA) \emph{(\bibinfo{series}{CHI '20})}. \bibinfo{publisher}{Association for Computing Machinery}, \bibinfo{address}{New York, NY, USA}, \bibinfo{pages}{1–14}.
\newblock
\showISBNx{9781450367080}
\href{https://doi.org/10.1145/3313831.3376516}{doi:\nolinkurl{10.1145/3313831.3376516}}


\end{thebibliography}

\appendix









\end{document}